\documentstyle[11pt,newpasp,twoside,epsfig]{article}
\def\edcomment#1{\iffalse\marginpar{\raggedright\sl#1\/}\else\relax\fi}
\reversemarginpar

\begin{document}
\title{Eigenspectra of the SDSS DR1 quasars}
\author{Ching-Wa Yip, Andrew J. Connolly, \& Daniel E. Vanden Berk}
\affil{Department of Physics and Astronomy, University of Pittsburgh,
Pittsburgh, PA 15260}
\author{Zhaoming Ma, Joshua A. Frieman, \& Mark SubbaRao}
\affil{Department of Physics and Astrophysics, University of Chicago, Chicago, IL 60637}
\author{Alex S. Szalay}
\affil{Department of Physics and Astronomy, Johns Hopkins University, Baltimore, MD 21218}

\begin{abstract}
We construct eigenspectra from the DR1 quasars in the 
SDSS using the Karhunen-Lo\`eve (KL) transform (or Principal Component Analysis, PCA) in different
redshift and luminosity bins. We find that the quasar spectra can be classified, by the
first two eigenspectra, into a continuous sequence in the variation of the spectral slope. We also
find a dependence on redshift and luminosity in the eigencoefficients. The
dominant redshift effect is the evolution of the blended Fe~II emission (optical) and the Balmer continuum
(the ``small-bump'', $\lambda_{rest} \approx 2000-4000$\AA), while the luminosity effect is related to 
the Baldwin effect. Correlations among several major broad emission lines
are found, including the well-known ``Eigenvector-1''. 
\end{abstract}

\section{Why Study Quasar Eigenspectra?}

Spectroscopic observations from the 
Sloan Digital Sky Survey (SDSS) have the advantage of large numbers of quasars (QSOs) and a 
large redshift range that provides an unique opportunity to study their intrinsic properties, 
sample variation, and spectral classification in great detail. 
A powerful method to address these questions is the
KL transform. The idea is to derive from the observed spectra
a lower dimensional set of eigenspectra (Connolly et al.\ 1995), in which
the essential spectral properties are represented. In the following
 we outline some results from applying this method to the SDSS quasars.

\section{Data and KL Transform}

The sample we use is the Sloan Digital Sky Survey (SDSS; York et al.\ 2000) First
Data Release (DR1; Abazajian et al.\ 2003) quasar catalog (Schneider et al.\ 2003),
consisting of 16,713 quasars. The redshifts range from 0.08 to 5.41 and the K-corrected $i$-band 
absolute 
magnitudes ($M_i$) from -30 to -22. We divide the whole sample into several redshift and luminosity 
bins with $\Delta z$ ranging from $\approx 0.45$ to $1.8$ and $\Delta M_i = 2$.
We consider wavelength ranges for all bins within a restframe 900\,\AA\ to 8000\,\AA. 
The spectra are shifted to their restframes, and bad data 
are approximated using the gap-repairing formalism (Connolly \& Szalay 1999). 
The details of the KL transform are described in, for example, Connolly et al.\ (1995).
Previous work applying the KL transform on quasar spectra or observed parameters 
include Francis et al.\ (1992), Boroson \& Green (1992), and Shang et al.\ (2003).
In the following, the flux densities are expressed in restframe values and the wavelengths are in vacuum.

\begin{figure}
\begin{center}
\plotfiddle{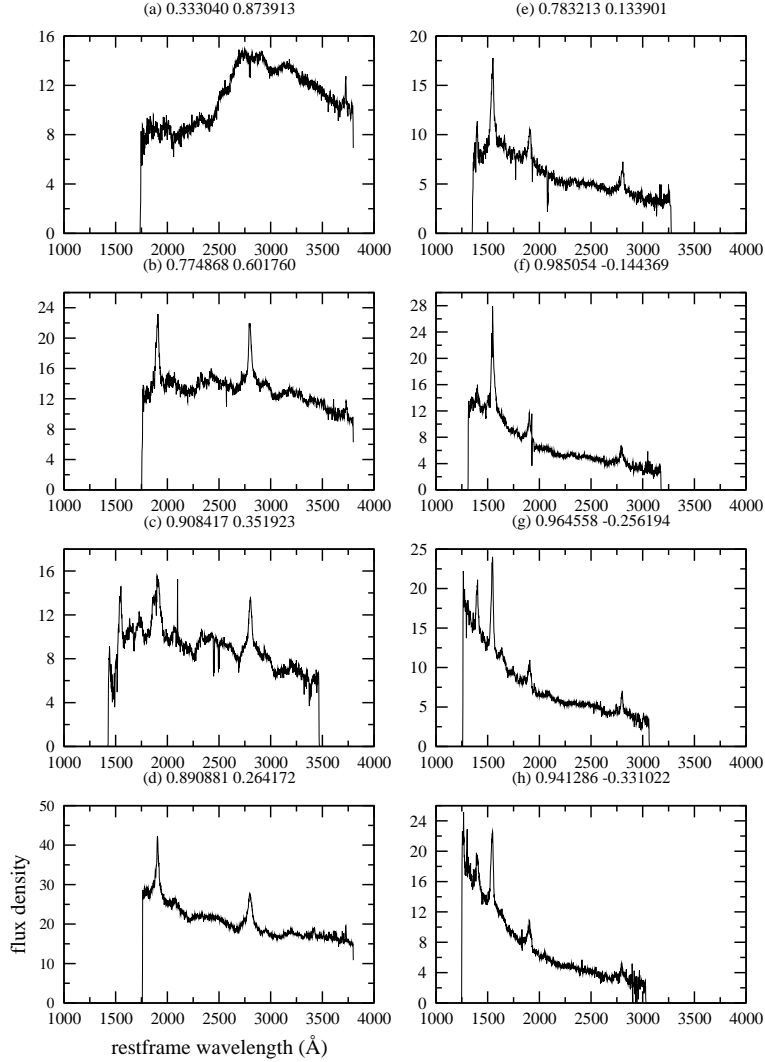}{5.5in}{0.0}{60}{60}{-175}{-40}
\caption{Observed quasar spectra for different values of the first and second eigencoefficients 
(in a particular redshift and luminosity bin).} 
\end{center}
\vspace{-2.0cm}
\end{figure}

\section{Results}

We find that the quasars can be classified into a continuous variation of the spectral slope. Figure~1 
shows the observed quasar spectra as a function of the first two eigencoefficients, $a_1$ 
and $a_2$ (in our notation the first mode is the mean spectrum), in the redshift range $1.16-2.06$
and the luminosity range $-28$ to $-26$. The actual values of the eigencoefficients 
are shown on top of each sub-figure. Along the sequence with decreasing $a_2$ values, the quasar continua 
are progressively bluer. This implies that the linear-combination of the first 2 modes changes the spectral 
slope. In this regard, this is similar to the galaxy spectral classification by the KL approach
(Connolly et al.\ 1995).

\begin{figure}
\plotfiddle{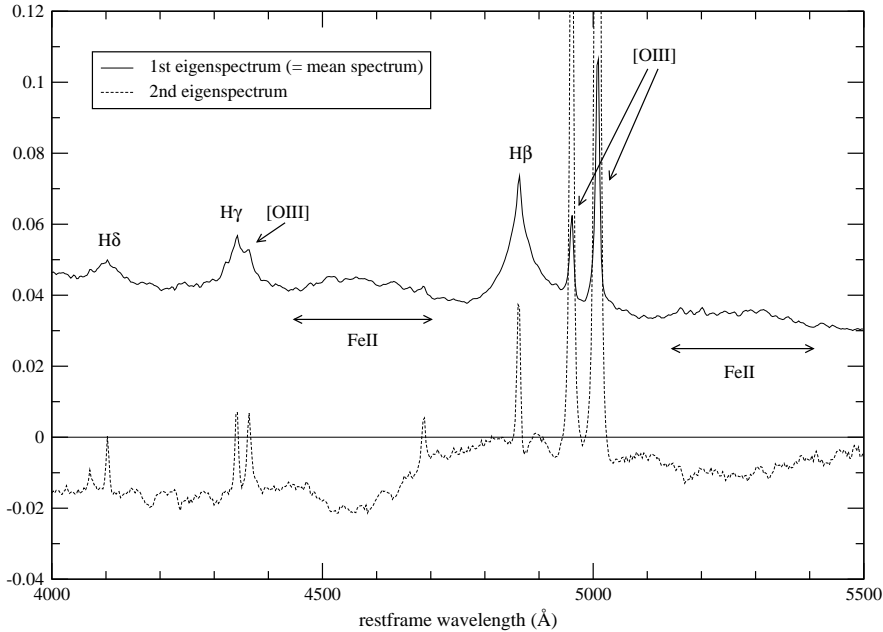}{3.5in}{0.0}{50}{50}{-190}{-30}
\caption{Eigenspectra of the SDSS DR1 quasars around the H$\beta$ region.}
\end{figure}

Figure~2 shows the first two eigenspectra in the local region around H~$\beta$ 
($\lambda_{rest} = 4000-5500$\AA), constructed with more than 2,500 SDSS QSOs in the redshift
range $0.08 - 0.67$. An anti-correlation is seen between Fe~II (optical) and [O~III]$\lambda$5008, 
which is also known as the ``Eigenvector-1'' (Boroson \& Green, 1992). It is generally believed that
the [O~III]$\lambda$5008 luminosity is an isotropic property, as such this anti-correlation is not driven
by the observed orientation of the quasar.

In the local region around Ly~$\alpha$ ($\lambda_{rest} = 1150-2000$\AA), the eigenspectra
from about 1300 wavelength-selected SDSS QSOs show agreement with those from Francis et al.\ (1992).
The second eigenspectrum shows low-velocity (so-called {\it line-core}) components of 
broad emission lines, and the third mode gives the continuum slope variation.

We find a dependence of the coefficient from the second eigenspectrum on both redshift and
luminosity (not shown). The dominant redshift effect is found to be 
the evolution of the blended Fe~II emission lines (optical) and the Balmer continuum (the ``small-bump'', 
$\lambda_{rest} \approx 2000-4000$\AA), as such it is
more prominent in the lower-redshift quasars and less obvious or absent in the higher-redshift ones. This
result suggests that the Iron abundances build up as quasar evolves, 
in agreement with quasars in the Large, Bright Quasar Survey (Green et al.\ 2001)
and in another independent study (Kuhn et al.\ 2001). On the other hand,
the dominant luminosity effect in the quasar spectra is found to be the Baldwin effect --- the
decrease of line equivalent width with luminosity,
which is detected in Ly~$\alpha$, Si~IV+O~IV$]$, C~IV, He~II, C~III$]$ and Mg~II emission lines.
The effect in N~V seems to be redshift dependent, so that an opposite sign (an
{\it increase} of line equivalent width with luminosity) is seen in the lower-redshift quasars.

In summary, we show that the quasar eigenspectra are invaluable to us in probing 
the intrinsic properties of quasars in a large data set. In addition to
the redshift and luminosity dependent sets of eigenspectra, we also
construct a single set of eigenspectra of the full DR1 catalog ranging from
restframe wavelengths 900\AA \, to 8000\AA \, in order to study 
the correlations between the UV and the optical spectral regions. Interested readers can 
find detail discussions in properties of these eigenspectra and future directions in Yip et al.\ (2003). 


Funding for the creation and distribution of the SDSS Archive has been
provided by the Alfred P. Sloan Foundation, the Participating
Institutions, the National Aeronautics and Space Administration, the
National Science Foundation, the U.S. Department of Energy, the
Japanese Monbukagakusho, and the Max Planck Society. The SDSS Web site
is http://www.sdss.org/.

\end{document}